%% file: main.tex
\documentclass[fleqn,twoside,twocolumn,nofootinbib,showkeys,aps,numerical,floatfix]{revtex4}
\usepackage[sec,nocpr]{ujp}

\input{preamble.tex}

\begin{document}
	\title[Phenomenology of multiplicity distributions]%
	{PHENOMENOLOGY OF CHARGED-PARTICLE MULTIPLICITY DISTRIBUTIONS (to appear in Ukr.~J.~Phys.)}%
	\author{Anton Alkin}%
	\affiliation{Bogolyubov Institute for Theoretical Physics, Nat. Acad. of Sci. of Ukraine}%
	\thanks{14b, Metrolohichna Str., Kyiv 03143, Ukraine}%
	\email{alkin@bitp.kiev.ua}%
	\udk{539} \pacs{No. 13.85.Hd} \razd{\seci}%
	\autorcol{A.~Alkin}%
	\setcounter{page}{1}%
	\keywords{charged-particle multiplicity, proton-proton scattering, LHC}
	
\begin{abstract}
Charged-particle multiplicity distributions are an interesting tool to study both soft- and hard-QCD processes in hadronic collisions. Since last century a significant range of center-of-mass energies has been probed, ranging from a few \GeV\xspace to 13 \TeV\xspace in the latest LHC run. Common analysis of multiplicity distributions at different energies, in different phase space regions and from sufficiently different experiments provides a way to systematize and review existing phenomenological models of multiple particle production. In this work a phenomenological model is suggested, that can describe simultaneously charged-particle multiplicity distributions in different restricted pseudorapidity intervals for proton-proton collisions. The model is successfully applied to experimental results of ALICE experiment at LHC. 
\end{abstract}%

\maketitle

\section{Introduction}
\label{sec:intro}%
\input{introduction.tex}%

\section{Multiplicity distributions in collider experiments}
\label{sec:experimental-multiplicity}%
\input{experimental-multiplicity.tex}%

\section{Models of multiplicity distributions}
\label{sec:multiplicity-models.tex}%
\input{multiplicity-models.tex}

\section{Conclusion}
\label{sec:conclusion}%
\input{conclusion.tex}%

\section*{Acknowledgments}
The author wishes to express his gratitude to G. M. Zinovjev, Ye. S. Martynov and Jean-Pierre Revol for their support and fruitful discussions.

This study is performed in a framework of targeted research program of Department of Physics and Astronomy, National Academy of Sciences of Ukraine.

\bibliographystyle{apsrev-improved}
\bibliography{bib}

\rezume{Алькiн Антон}{ФЕНОМЕНОЛОГIЯ РОЗПОДIЛIВ МНОЖИННОСТI ЗАРЯДЖЕНИХ ЧАСТИНОК}{Розподiли множинностi заряджених частинок є цiкавим iнструментом для вивчення як м’яких, так й жорстких КХД-процесiв у адронному розсiяннi. 
З попереднього сторiччя значний дiапазон енергiй зiткнення було експериментально розглянуто, вiд декiлькох ГеВ до 13 ТеВ в останньому перiодi роботи LHC.
Сукупний аналiз розподiлiв множинностi в зіткненнях протонів на рiзних енергiях, в рiзних регiонах фазового простору народжених частинок та вiд рiзних експериментальних колаборацiй дає можливiсть систематизувати та перевiрити наявнi феноменологiчнi моделi множинного народження частинок. 
В цiй роботи запропоновано феноменологiчну модель, що дозволяє одночасно описати розподiли множинностi заряджених частинок у зіткненнях протонів в рiзних iнтервалах псевдохуткостi. 
Модель успiшно застосовано до експериментальних розподiлiв множинностi, що отриманi колаборацiєю ALICE на LHC.

К л ю ч о в і\ \ с л о в а: множинність заряджених частинок, розсіяння протонів, LHC}
\end{document}

%% file: preamble.tex
\usepackage{hepunits}
\usepackage{amsmath}
\usepackage{amssymb}
\usepackage{amsfonts}
\usepackage[colorlinks=false,breaklinks=true]{hyperref}
\usepackage{graphicx}
\graphicspath{{figures-color/}}
\usepackage[caption=false]{subfig}
\usepackage{multirow,booktabs,dcolumn}
\usepackage{xspace,indentfirst,paralist}
\usepackage{csquotes}

\usepackage[capitalize]{cleveref}

\newcommand{\pt}{\ensuremath{p_{\bot}}}
\newcommand{\avg}[1]{\ensuremath{\langle #1\rangle}}
\newcommand{\slfrac}[2]{\left.#1\right/#2}

%% file: introduction.tex
Charged particle multiplicity distributions are basic and general observables in modern collider experiments, by their nature containing information both on soft QCD processes (that still dominate even at LHC energies) and hard scattering thus allowing to explore both components and their interrelation. 
As a direct measure of a collision event inelasticity, multiplicity potentially contains information on the various features of particle production mechanisms and hadronization process. 
However, while in heavy ion physics there is a direct correspondence between multiplicity and centrality \cite{Miller:2007ri}, the situation in proton-proton scattering is much less clear. 
With the recent unambiguous information on collectivity in pp scattering at LHC energies \cite{ALICE:2017jyt}, as well as potentially non-trivial transverse structure of interaction region \cite{Alkin:2014rfa}, the possible relation between multiplicity and the underlying interaction becomes even more interesting.
However, as usual when dealing with observables of statistical nature, we must always bear in mind that it can be largely independent of the underlying dynamical process. 
In this paper we will explore a straightforward phenomenological model of charged-particle multiplicity distributions in restricted phase space, that allows disentangling instrumental limitation of the collider experiments to extract a full phase space charged-particle multiplicity distribution and study its behavior.

%% file: experimental-multiplicity.tex
\subsection{Pre-LHC era}
Both experimental multiplicity measurements and the theoretical attempts to describe them have a rich history. 
Detailed review of contemporary experimental data theoretical understanding was made by Carruthers and Shih in 1987 \cite{Carruthers:1987tv}. 
Here we have not repeated the analysis of non-pp multiplicities and old pp data. 
While combined analysis of the very wide center-of-mass energy range probed for a few decades of collider experiments may yield interesting results, the limited precision and multiplicity range of older data would not contribute in a significant way.
This also means that available data wasn't useful in distinguishing and excluding models. 
This remained mostly true up until the center-of-mass energy of 900 \GeV\xspace was reached in UA5 experiment and the presence of a possible structure in multiplicity distribution was identified \cite{Fuglesang:1989st}.
It is important to emphasize, that this structure was found only in a full phase space multiplicity distribution at the time.
A similar observation was later made by ALICE collaboration at LHC \cite{aamodt2010charged09} for the same collision energy, but in a restricted phase space distribution.
The range of multiplicities in restricted phase space distributions was extended compared to that of UA5, thus indicating that the structure is already present in central pseudorapidity window at 900 \GeV\xspace.
This will be considered in more detail in \cref{sec:multiplicity-structures}.
\subsection{Multiplicity in LHC experiments}
Since its start in 2008, LHC produced an enormous amount of data on proton-proton, proton-ion and ion-ion collisions at center-of-mass energies ranging from 0.9 to 13 \TeV\xspace with varying beam conditions.
Multiplicity distributions obtained in LHC experiments are thus of special interest as they are produced with largely unchanged experimental setups and measurement procedures, enabling a combined analysis in which systematic uncertainties can be controlled in an assumption that they are similar between the measurements at different energies. 
In an attempt to model restricted phase space multiplicity distributions, we concentrate on those measured in symmetric $\eta$ regions. 
From available LHC experiments, only CMS \cite{Khachatryan:2010nk} and ALICE \cite{Adam:2015gka} have published a large set of data in varying pseudorapidity ranges.
ALICE presented measurements at more energies, including 8~\TeV\xspace, for $|\eta|$ up to 1.5, and presented a preliminary result for multiplicity distribution measurement in wider ranges, up to $\left|\eta\right| < 3.4$ \cite{Zaccolo:2015udc}.
Additionally, the effort was made by ALICE collaboration to refine the non-single-diffractive event sample (i.e., inelastic events excluding single-diffractive events).
However, the wide pseudorapidity range data points are not yet available at the time of this publication.
\text{ATLAS} Collaboration measurements involve restriction on particle transverse momentum, modeling which is beyond the scope of this paper.
Measurements at $\sqrt{s} = 13$~\TeV\xspace aren't yet available as well.

It is important to discuss the event class normalization and the limitations that are inherent to restricted phase space multiplicity distributions. 
There is an obvious theoretical preference to dealing with non-single-diffractive event samples (NSD), as diffractive events are known to have properties that are somewhat different from the general inelastic events. 
However, due to the nature of these events, they have a tendency to have very low (or no at all) multiplicity in the central (pseudo)rapidity thus contributing mostly to value of $P(0)$ ($P(N)$ being the restricted phase space multiplicity distribution) and thus only affect the overall normalization.
Moreover, excluding single-diffractive events requires modeling their cross-section fraction and other properties within event generators used to simulate various detector inefficiencies therefore increasing systematic uncertainty of multiplicity distribution. 
As demonstrated by ALICE \cite{Adam:2015gka}, the total systematic uncertainties are strongly correlated bin-to-bin and cannot be disentangled in a simple way complicating phenomenological analysis of the experimental data.
In this paper we, however, use multiplicity distributions normalized to non-single-diffractive event sample, keeping in mind that any measurement procedure can reliably remove from multiplicity distribution only those diffractive events that the detector set is sensitive to.
In particular, ALICE only considers events with diffractive system mass $M_{X} \leq 200$~\GeV\xspace \cite{alice-crosssection-diffraction} as single-diffractive.
The residual single-diffractive events in the sample affect the low-multiplicity part of the multiplicity distribution.

As was already mentioned in the previous paragraph, the value $P(0)$ depends mostly on phase space restrictions. 
This value should be explicitly modeled as the effect of geometrical restriction on particle counting region accounting for the spatial distribution of produced particles. 
However, even without explicit model, this value can be used to put a restriction on scaling factor introduced by ALICE \cite{Adam:2015gka} in parametrization of the measured distributions.
We leave the details to be discussed within the next section.

%% file: multiplicity-models.tex
\subsection{Negative-binomial distribution}
A number of discrete probability distributions were used to model hadronic multiplicities. The most successful description (up to center-of-mass energy of 900~\GeV\xspace) was achieved with a single \emph{negative binomial} discrete probability distribution (NBD) \cite{Giovannini:1985mz} that we will define here as
\begin{multline}
\label{eq:nbd-definition}
P(n;\avg{n},k) = \frac{\Gamma(k+n)}{\Gamma(k)\Gamma(n+1)}\left[\frac{\avg{n}}{k+\avg{n}}\right]^n \times \\
 \times \left[\frac{k}{k+\avg{n}}\right]^{k}\!\!\!,
\end{multline}%
where $\avg{n}$ is the average value and $k$ is a shape parameter related to distribution variance $D$ by
\begin{equation}
\frac{D^2}{\avg{n}^2} = \frac{1}{\avg{n}} + \frac{1}{k}.
\end{equation}%
Note that setting $k=1$ yields \emph{Bose-Einstein} distribution, while in the limit $k\to\infty$ we get \emph{Poisson} distribution.

Almost half a century ago, first Polyakov \cite{polyakov1971similarity} and then Koba, Nielsen and Olesen (KNO) \cite{koba1972scaling} independently suggested that at high enough center-of-mass energy, the probability $P(N)$ of producing $N$ particles in a collision should reach an asymptotic shape, when expressed as a function of $z=\slfrac{N}{\avg{N}}$ 
\begin{equation}
P(N) = \frac{1}{\avg{N}}\psi(z).
\end{equation} 
While it was initially found to hold for a limited range of multiplicities already at ISR energies \cite{breakstone1984charged} for non-single-diffractive event sample, it was violated for all inelastic events.
It was initially found to hold for non-single-diffractive events at energies up to $\sqrt{s} = 900$~\GeV\xspace \cite{aamodt2010charged09}, however, further analysis \cite{Adam:2015gka} showed that it is significantly violated for LHC energies in all available $\eta$ intervals even for NSD events.
Using the KNO variable $z$ we can derive behavior of NBD in the scaling limit defined by
\begin{equation}
n\to\infty,\quad \avg{n}\to\infty, \quad \text{fixed}\ z,
\end{equation}%
as $\avg{n}/k \gg 1$, \cref{eq:nbd-definition} becomes
\begin{equation}
	P(n;k) = \frac{1}{\avg{n}}\psi(z;k)
\end{equation}
with
\begin{equation}
	\psi(z;k) \equiv \frac{k^k}{\Gamma(k)}z^{k-1}e^{-kz}
\end{equation}
that is a special case of \emph{gamma} distribution
\begin{equation}\label{eq:gamma-definition}
\psi(x;k,\gamma) = \frac{\gamma^k}{\Gamma(k)}x^{k-1}e^{-\gamma x}
\end{equation}
where the shape parameter $\gamma$ coincides with $k$ due to a requirement that $\avg{z} = 1$.
We will use gamma distribution rewritten with $\avg{x}$ as a parameter
\begin{equation}\label{eq:gamma-definition-with-avg}
\psi(x;k,\avg{x}) = \frac{k}{\avg{x}} \frac{1}{\Gamma(k)} \left(\frac{kx}{\avg{x}}\right)^{k-1} e^{-\frac{kx}{\avg{x}}}.
\end{equation}

Based on the properties described above we will consider a set of possible approximations for the restricted phase space multiplicity distributions using combinations of the functions defined above. 
\subsection{Structures in multiplicity distributions}
\label{sec:multiplicity-structures}
After the inability of a single negative binomial distribution to describe the shape of multiplicity distribution was discovered, a number of papers examined possible scenarios and their phenomenological consequences where multiplicity distribution consists of two incoherent components, designated \emph{soft} and \emph{semi-hard}, both described by NBDs, with the total distribution being a weighted sum \cite{Giovannini:1998zb,PhysRevD.60.074027}. 
These ideas were applied to CMS Collaboration non-single-diffractive multiplicity distributions by Gosh in 2012 \cite{bPremGosh}.
While none of the proposed scenarios was found to be realized, some important conclusions could be derived. 
The relative importance of a second (\enquote{semi-hard}) NBD term grows with center-of-mass energy as well as the size of pseudorapidity window, however it can be argued that the multiplicity distributions in the most restricted ranges can be suitably well described by a single NBD.
The first (\enquote{soft}) NBD term was found to be almost independent of $\sqrt{s}$ hinting on possible partial scaling behavior in charged-particle multiplicity distributions. 
Similar analysis, performed by ALICE Collaboration \cite{Adam:2015gka} confirmed some observations, however it also emphasized the particular issue with modeling distributions that are strongly correlated bin to bin due to a nature of the measurement. 
We slightly improve this analysis by introducing a reasonable constraint on the fits and consider other possibilities for both first and the second term of the weighted sum.
Finally, we consider a unified description of the restricted phase space distributions in different $\eta$ ranges for a fixed $\sqrt{s}$ and discuss the possible mechanisms of generating the second term as well as their phenomenological consequences.

It is worth mentioning, that another type of structures in experimental multiplicity distributions was identified by examining the recurrent relation representation of the probability distributions \cite{Wilka:2016ufh,Wilk:2016cnp}
\begin{equation}
(n+1)P(n+1)=\avg{n}\sum\limits_{i=0}^{n} C_i P(n-i).
\end{equation}
However, it should be noted, that experimental multiplicity distribution is obtained with regularized unfolding and thus contains residual oscillations with a \enquote{period} that is proportional to $\sqrt{n}$, as indicated by the structure of unfolding response matrix~\cite{Adam:2015gka}.
The observed property of recurrent relations coefficients $C_i$ to detect an oscillatory structure at low multiplicities is thus can be explained as an artifact of regularized unfolding \cite{cowan-stat-analysis}.
One of the ways to investigate the possibility of physical oscillating structures in multiplicity distributions would be to re-formulate unfolding problem in terms of recurrent relation coefficients and study their dependence on statistical fluctuations in the raw data.
\subsection{Modeling multiplicity in restricted phase space}
\subsubsection{Direct approach}
We will follow the original approach of ALICE \cite{Adam:2015gka}, that introduces a scaling factor $\lambda$ in a way that
\begin{equation}\label{eq:0-bin-constraint}
\lambda\sum\limits_{n=1}^{n_{\mathrm{max}}} P(n;\mathbf{p}) = \sum\limits_{n=1}^{n_{\mathrm{max}}} P_{\mathrm{exp}}(n),
\end{equation}
i.e. the model probability distribution $P(n;\mathbf{p})$, with its best-fit parameters $\mathbf{p}$, is re-normalized to be compatible with the experimental distribution in the available range of multiplicities. 
This is needed to account for the fact that, in the model distributions used, values at $n=0$ are lower than at $n=1,2,\ldots$ up to the distribution peak, while in the experimental distributions $P_{\mathrm{exp}}(0)$ is larger than the rest of the distribution due to phase space restrictions and thus the reminder of the distribution is scaled down.
Noting, that the overall normalization of the experimental distribution is dependent on the value of $P_{\mathrm{exp}}(0)$, we can add a constraint for the fit criterion based on this value. Using \cref{eq:0-bin-constraint} and the fact that both experimental and model distributions are normalized to 1, we can write
\begin{gather}
P_{\mathrm{exp}}(0) + \sum\limits_{n=1}^{n_{\mathrm{max}}} P_{\mathrm{exp}}(n) + \Delta\left(\sum P_{\mathrm{exp}}\right) = 1,\\ 
\Delta\left(\sum P_{\mathrm{exp}}\right) \approx  0\\
P_{\mathrm{exp}}(0) + \lambda\sum\limits_{n=1}^{n_{\mathrm{max}}} P(n;\mathbf{p}) \approx 1, \\
\Delta\chi^2 = \left( \frac{ P_{\mathrm{exp}}(0) + \lambda\sum\limits_{n=1}^{n_{\mathrm{max}}} P(n;\mathbf{p}) - 1 }{\sigma_0} \right)^2\!\!\!,
\end{gather} 
where $\sigma_0$ is the uncertainty on $P_{\mathrm{exp}}(0)$ and we are neglecting the reminder of the distribution sum for $n > n_{\mathrm{max}}$ as multiplicity distributions decrease very fast.
The full fit criterion therefore is written as
\begin{multline}
\chi^2 = \beta \Delta\chi^2\left(P_{\mathrm{exp}}(0), P(n;\mathbf{p})\right) + \\ 
+ \sum\limits_{n=1}^{n_{\mathrm{max}}} \left( \frac{\lambda P\left(n;\mathbf{p}\right) - P_{\mathrm{exp}}(n)}{\sigma_n} \right)^2\!\!\!,
\end{multline}
where we introduce Lagrange multiplier $\beta$ that will be adjusted to improve fit quality but will not be considered a free parameter.
An obvious function of the factor $\beta$ is to off-set the relatively large $\sigma_0$ uncertainty and thus partially compensate for the distribution freedom within the correlation corridor.
It was found that the stable convergence is achieved at values ${\beta \sim 10^2}$.
\begin{table}[t]
	\begin{ruledtabular}
		\centering
		\caption{\label{tab:fit-options}Functional forms of the weighted sum components as per \cref{eq:weighted-sum} with the typical $\chi^2$ values. Note that due to the significant correlated uncertainties in experimental distributions these values are at least order of magnitude lower than one would normally expect.}
		\vspace{1em}
		\begin{tabular}{rllc}
			{}	& \textbf{$P_1$}				& \textbf{$P_2$}				& \textbf{typical $\slfrac{\chi^2}{\mathrm{NDF}}$} \\
			\midrule
			1	&	{NBD}			& {NBD}				& 0.1 \\
			2	&	{Poisson}		& {NBD}				& 10 \\
			3	&	{NBD}			& {Bose-Einstein}	& 10 \\
			4	&	{NBD}			& {Gamma}			& 0.1 \\
		\end{tabular}%
	\end{ruledtabular}
\end{table}%
\begin{table*}[t]
	\begin{ruledtabular}
	\centering
	\caption{\label{tab:best-fit}Best-fit parameter values for the NBD$+$NBD and NBD$+$Gamma model distributions for all $\sqrt{s}$ and $\eta$ ranges considered. Note that parameters uncertainties are calculated assuming uncorrelated errors in experimental distribution and thus are unreliable.}
	\vspace{1em}
	\begin{tabular}{>{\bfseries}ccccccccc}
		\multicolumn{9}{c}{\textbf{NBD+NBD}} \\
		\midrule[2pt]
		$\sqrt{s}$ (\TeV\xspace)		& \textbf{$\left|\eta\right|<$}	& \textbf{$\lambda$}	& \textbf{$\alpha$}	& \textbf{$\avg{n}_1$}	& \textbf{$k_1$}	& \textbf{$\avg{n}_2$}	& \textbf{$k_2$} & \textbf{$\slfrac{\chi^2}{\mathrm{NDF}}$} \\
		\midrule
		\multirow{3}{*}{0.9}	& 0.5	& $0.93\pm 0.06$	& $0.43\pm 0.53$	& $2.1\pm 1.9$	& $3.3\pm 13.9$	& $5\pm 4$	& $2.8\pm 3.4$	& 0.167 / 30 \\
		{}						& 1		& $0.94\pm 0.02$	& $0.58\pm 0.25$	& $5.0\pm 2.4$	& $2.9\pm 1.7$	& $13\pm 6$	& $3.8\pm 3.0$	& 0.162 / 54 \\
		{}						& 1.5	& $0.96\pm 0.01$	& $0.76\pm 0.28$	& $9.0\pm 5.1$	& $2.4\pm 1.1$	& $22\pm 14$	& $5.2\pm 5.6$	& 3.415 / 66 \\
		\midrule
		\multirow{3}{*}{2.76}	& 0.5	& $0.93\pm 0.03$	& $0.50\pm 0.16$	& $2.5\pm 1.0$	& $2.8\pm 2.4$	& $7\pm 2$	& $3.0\pm 1.2$	& 0.421 / 44 \\
		{}						& 1		& $0.94\pm 0.01$	& $0.55\pm 0.09$	& $5.3\pm 1.2$	& $2.7\pm 0.9$	& $16\pm 2$	& $3.5\pm 0.9$	& 0.403 / 77 \\
		{}						& 1.5	& $0.95\pm 0.01$	& $0.68\pm 0.15$	& $9.8\pm 3.6$	& $2.2\pm 0.6$	& $27\pm 8$	& $4.4\pm 2.4$	& 6.598 / 99 \\
		\midrule
		\multirow{3}{*}{7}		& 0.5	& $0.94\pm 0.02$	& $0.70\pm 0.12$	& $3.6\pm 1.4$	& $1.8\pm 0.7$	& $12\pm 3$	& $4.1\pm 1.6$	& 0.874 / 62 \\
		{}						& 1		& $0.94\pm 0.01$	& $0.66\pm 0.05$	& $7.0\pm 1.1$	& $2.0\pm 0.4$	& $23\pm 2$	& $4.2\pm 0.7$	& 1.871 / 110 \\
		{}						& 1.5	& $0.95\pm 0.01$	& $0.60\pm 0.04$	& $9.9\pm 1.2$	& $2.1\pm 0.3$	& $32\pm 3$	& $3.7\pm 0.5$	& 8.032 / 146 \\
		\midrule
		\multirow{3}{*}{8}		& 0.5	& $0.93\pm 0.02$	& $0.57\pm 0.10$	& $3.1\pm 1.0$	& $2.0\pm 1.0$	& $11\pm 2$	& $3.2\pm 1.0$	& 0.837 / 60 \\
		{}						& 1		& $0.93\pm 0.01$	& $0.61\pm 0.05$	& $6.6\pm 1.2$	& $2.1\pm 0.4$	& $22\pm 2$	& $3.8\pm 0.7$	& 1.491 / 106 \\
		{}						& 1.5	& $0.94\pm 0.01$	& $0.70\pm 0.06$	& $11.6\pm 2.2$	& $1.8\pm 0.2$	& $37\pm 5$	& $4.6\pm 1.1$	& 5.976 / 138 \\
		\midrule
		\multicolumn{9}{c}{\textbf{NBD+Gamma}} \\
		\midrule[2pt]
			$\sqrt{s}$ (\TeV\xspace)		& \textbf{$\left|\eta\right|<$}	& \textbf{$\lambda$}	& \textbf{$\alpha$}	& \textbf{$\avg{n}_1$}	& \textbf{$k_1$}	& \textbf{$k_2$}	& \textbf{$\avg{n}_2$} & \textbf{$\slfrac{\chi^2}{\mathrm{NDF}}$} \\
		\midrule
		\multirow{3}{*}{0.9}	& 0.5	& $0.62\pm 0.05$	& $0.80\pm 0.02$	& $5.6\pm 0.4$	& $2.96\pm 0.45$	& $2\pm 1$	& $2.6\pm 0.4$	& 0.155 / 30 \\
		{}						& 1		& $0.84\pm 0.20$	& $0.70\pm 0.20$	& $5.0\pm 2.8$	& $2.91\pm 1.53$	& $4\pm 3$	& $13.7\pm 6.8$	& 0.162 / 54 \\
		{}						& 1.5	& $0.88\pm 0.19$	& $0.89\pm 0.09$	& $9.5\pm 4.1$	& $2.34\pm 0.95$	& $5\pm 4$	& $24.7\pm 12.0$	& 3.404 / 66 \\
		\midrule
		\multirow{3}{*}{2.76}	& 0.5	& $0.98\pm 0.06$	& $0.55\pm 0.15$	& $2.5\pm 1.1$	& $2.66\pm 1.84$	& $3\pm 1$	& $8.3\pm 2.2$	& 0.414 / 44 \\
		{}						& 1		& $0.79\pm 0.07$	& $0.70\pm 0.07$	& $5.4\pm 1.3$	& $2.65\pm 0.80$	& $3\pm 1$	& $16.7\pm 2.7$	& 0.407 / 77 \\
		{}						& 1.5	& $0.80\pm 0.22$	& $0.86\pm 0.10$	& $10.1\pm 4.9$	& $2.18\pm 0.76$	& $4\pm 3$	& $28.1\pm 11.5$	& 6.594 / 99 \\
		\midrule
		\multirow{3}{*}{7}		& 0.5	& $0.94\pm 0.02$	& $0.76\pm 0.12$	& $3.9\pm 1.7$	& $1.74\pm 0.74$	& $4\pm 2$	& $12.9\pm 3.8$	& 0.877 / 62 \\
		{}						& 1		& $0.82\pm 0.04$	& $0.81\pm 0.04$	& $7.3\pm 1.2$	& $1.95\pm 0.36$	& $4\pm 1$	& $24.0\pm 2.5$	& 1.915 / 110 \\
		{}						& 1.5	& $0.72\pm 0.05$	& $0.82\pm 0.03$	& $10.0\pm 1.3$	& $2.10\pm 0.30$	& $4\pm 1$	& $32.5\pm 2.8$	& 8.012 / 146 \\
		\midrule
		\multirow{3}{*}{8}		& 0.5	& $0.92\pm 0.03$	& $0.64\pm 0.11$	& $3.2\pm 1.1$	& $1.98\pm 0.89$	& $3\pm 1$	& $11.5\pm 2.3$	& 0.830 / 60 \\
		{}						& 1		& $0.77\pm 0.05$	& $0.77\pm 0.04$	& $6.8\pm 1.3$	& $2.12\pm 0.42$	& $4\pm 1$	& $23.4\pm 2.6$	& 1.538 / 106 \\
		{}						& 1.5	& $0.77\pm 0.07$	& $0.87\pm 0.03$	& $11.9\pm 2.2$	& $1.80\pm 0.24$	& $4\pm 1$	& $37.9\pm 4.9$	& 5.974 / 138 \\
	\end{tabular}%
	\end{ruledtabular}
\end{table*}%

Consider a generic functional form of multiplicity distribution defined as a weighted sum of two probability distributions $P_1$, $P_2$ with their respective parameter sets $\mathbf{p}_1$, $\mathbf{p}_2$
\begin{multline}\label{eq:weighted-sum}
P(n;\mathbf{p}) = \alpha P_1(n;\mathbf{p}_1) + (1-\alpha)P_2(n;\mathbf{p}_2),\\ 0 < \alpha < 1.
\end{multline}
The combinations of probability distributions, used in this analysis, are presented in \cref{tab:fit-options} together with their typical $\chi^2$ values. 
Note that Bose-Einstein distribution can be only used as a second term as it lacks a distinct peak that is present in the experimental multiplicity distribution at low $n$, while Poisson distribution is too narrow to be used as a second term.
Gamma distribution is used in its two-parameter form (\cref{eq:gamma-definition-with-avg}).
Only the sum of two NBDs and the sum of NBD and gamma distribution are in agreement with data, providing a similar description. 
The best-fit parameters for these two cases are presented in \cref{tab:best-fit} for all the $\sqrt{s}$ and $\eta$ ranges considered. 
A trend can be identified in $\chi^2$ values, namely the apparent reduction in fit quality with extending $\eta$ window.
This can be attributed to the fact that the fraction of residual single- and normal double-diffractive events, that have non-zero multiplicity, in a given pseudorapidity region increases with its size changing the behavior of multiplicity distribution at low $n$, where the contribution to $\chi^2$ is the largest.
It is important to emphasize here, that while the parameter values themselves do provide a reasonable description of experimental curves, the uncertainties should be considered with great caution due to correlations within systematic uncertainties of experimental curves. 
While best-fit parameters for the sum of two NBDs differ slightly from those reported by ALICE~\cite{Adam:2015gka}, the curves they define lie within the correlation corridors provided with the experimental curves.

As expected, no clear trend is observed in parameter evolution, both as the function of energy and pseudorapidity window.
This is consistent with earlier analysis~\cite{bPremGosh} and confirms that overall multiplicity distribution cannot be unambiguously separated into different event classes with different multiple particle production mechanisms based on the distribution shape alone.
This is due to a significant overlap in multiplicity ranges between the two components, which increase the range of weight $\alpha$ variation.
However, a simultaneous description of multiplicity distributions in several pseudorapidity ranges is expected to be more reliable for this purpose.

The quality of description with weighted sum of NBD and Gamma distribution is on par with that of two NBDs. 
We note that, within uncertainties, the parameter $k \approx 4$ of the Gamma term seems to be independent of center-of-mass energy which is consistent with a scaling limit for the second component reached already at $\sqrt{s} = 0.9$~\TeV. 
More precise measurements of multiplicity distributions are required to test this observation.
\subsubsection{Unified modeling}%
\paragraph{Multiplicity reduction model.}%
As was already mentioned in the discussion above, the values of reduced multiplicities in restricted phase space are determined by interplay between the restriction itself and the underlying spatial distribution of the particles, produced in a collision. 
Let us consider this effect in more detail. 
Let us assume there is an unrestricted multiplicity distribution $P_{\mathrm{tot}}(N)$, where $N$ is the total number of charged particles produced in an inelastic collision. 
Let $P(n|N,\Delta\eta)$ denote the conditional probability that $n$ charged particles out of total $N$ fit into a given $\Delta\eta$. 
We can calculate the average number of charged particles in the $\Delta\eta$ region when total $N$ is fixed as
\begin{equation}
\avg{n}_{\Delta\eta,N} = \sum\limits_{n \leq N} n P(n|N,\Delta\eta),
\end{equation}
and the overall average as
\begin{multline}\label{eq:avg-n-def}
\avg{n}_{\Delta\eta} = \sum\limits_{N} \avg{n}_{\Delta\eta,N} P_{\mathrm{tot}}(N) = \\ 
= \sum\limits_{N} \sum\limits_{n \leq N} n P(n|N,\Delta\eta) P_{\mathrm{tot}}(N).
\end{multline}
Spanning $P(n|N,\Delta\eta)$ in such a way that ${P(n > N |N,\Delta\eta) \equiv 0}$, we can change the order of summation in \cref{eq:avg-n-def} so it becomes
\begin{equation}
\avg{n}_{\Delta\eta} = \sum\limits_{n} n \left( \sum\limits_{N} P(n|N,\Delta\eta) P_{\mathrm{tot}}(N) \right)\!\!,
\end{equation}
and thus the quantity
\begin{equation}\label{eq:restricted-Pn-def}
P_{\mathrm{mod}}(n,\Delta\eta) \equiv \sum\limits_{N} P(n|N,\Delta\eta) P_{\mathrm{tot}}(N)
\end{equation}
is the model for the restricted phase space multiplicity distribution we are interested in. 

It is important to emphasize here that to get the structure, similar to one in the observed distribution \cite{Adam:2015gka,Zaccolo:2015udc}, using the \cref{eq:restricted-Pn-def}, one of the compound distribution components, either total or conditional probability distributions, has to contain a similar structure. 
As we know that characteristic feature of multiplicity distribution is present for full phase space and becomes less apparent with decreasing pseudorapidity window, we will assume that $P_{\mathrm{tot}}(N)$ is its source.
It should be noted, that the hypothetical production mechanisms, that correspond to the event subsamples identified through multiplicity distribution, most likely differ both in the amount of particles produced in a typical act and the spatial configuration of the produced system and thus the factorization assumed in \cref{eq:avg-n-def} does not hold.
As an initial approximation we will, however, continue operating under an assumption that the hypothetical additional particle production mechanism generates the same spatial distribution on average.

In order for the \cref{eq:restricted-Pn-def} to be useful we must make some assumptions about the form of $P(n|N,\Delta\eta)$. 
We need to define its shape with a requirement that $n \leq N$ and that average observed multiplicity is a function of $N$, $\avg{n} = F(N,\Delta\eta)$.
In a borderline case of unrestricted $\eta$ range this distribution should collapse into an infinitely thin peak at ${\avg{n} = N}$. 
Assuming the shape of $\left. \slfrac{1}{n_{\mathrm{events}}}\slfrac{\mathrm{d}n_{\mathrm{particles}}}{\mathrm{d}\eta}\right|_{N}$ does not depend on $N$\footnote{this can be experimentally verified in the future, however there is no multiplicity-binned measurement of pseudorapidity density available for proton-proton scattering at the time of publication} we can start with a linear relation 
\begin{equation}\label{eq:A-meaning}
\avg{n} = A\left(\Delta\eta\right) N,
\end{equation}
where the constant parameter $A$ has simple meaning of the area fraction under pseudorapidity density curve within $\Delta\eta$ interval, that in a chosen approximation only depends on the size and position of $\Delta\eta$ interval.
A straightforward choice of $P(n|N)$ shape is thus a binomial distribution
\begin{equation}
\label{eq:binomial-restriction}
P(n|N) = \frac{N!}{n! \left( N-n\right)!} p^{n} \left(1 - p \right)^{(N-n)}
\end{equation}\label{eq:PnN-mod}
with mean $\avg{n} = Np$ and variance $D = Np(1-p)$ that identifies $p \equiv A$.

\paragraph{Full phase space multiplicity distribution.}%
With the explicit form of $P(n|N,\Delta\eta)$ in place, we can fit one model of $P_{\mathrm{tot}}(N)$ to a set of measurements in different pseudorapidity intervals at a given center-of-mass energy $\sqrt{s}$.

The choice of binomial restriction probability distribution (\cref{eq:binomial-restriction}), in particular, is convenient if the full distribution $P_{\mathrm{tot}}(N)$ is a NBD or a sum of NBDs.
Indeed, substituting \cref{eq:nbd-definition} into \cref{eq:restricted-Pn-def} we get
\begin{multline}
P(n) = \sum\limits_{N\geq n} {N \choose n} p^n (1-p)^{N-n} \times \\
\times {{N+k-1} \choose N} q^N (1-q)^k
\end{multline}
where we let 
\begin{equation}
q \equiv \frac{\avg{N}}{k+\avg{N}}.
\end{equation} 
This series, in fact, converges to NBD
\begin{equation}
P(n) = {{n+k-1} \choose n} z^n (1-z)^k
\end{equation}
with the same parameter $k$ and
\begin{equation}
z \equiv \frac{q p}{1-q + qp},
\end{equation}
from which we recover $\avg{n} = p\avg{N}$. 
An obvious conclusion is that the apparent double-NBD behavior of restricted phase space multiplicity distributions in this framework follows directly from the similar structure of the unrestricted distribution.
It should be noted here that above derivation is done for an integer $k$, however, can be shown to hold for real-valued $k$ as well. 

The considerations described above allow us to draw some immediate conclusions. 
From the ratios of average multiplicities in ALICE measurement \cite{Adam:2015gka} (see \cref{tab:average-nch-ratios}), we can estimate the expected ratios\footnote{$A_{1,2,3}$ is a shorthand for $A_{-0.5}^{+0.5}$, $A_{-1}^{+1}$ and $A_{-1.5}^{+1.5}$} $A_1$ : $A_2$ : $A_3$.
The reduction procedure leaves parameter (or parameters, in case of a weighted sum) $k$ of the original NBD intact. 
The shape parameter $k$ within the clan model \cite{Giovannini:1985mz} is related to an average number of particle clans
\begin{equation}
N_{\mathrm{clans}} = k \times \ln \left( 1 + \frac{\avg{n}}{k} \right),
\end{equation}
with the average number of particles per clan
\begin{equation}
n_{\mathrm{particles}} = \frac{\avg{n}}{N_{\mathrm{clans}}},
\end{equation}
and thus is expected to be determined solely by the underlying production mechanism and not the size of pseudorapidity window.
On the other hand, $k$ can be related directly to the two-particle correlation function \cite{Giovannini:1985mz}.
Ability (or inability) to describe all restricted phase space multiplicity distributions with one set of shape parameters (for a given weighted sum unrestricted multiplicity distribution model) is thus an important clue into the average spatial configuration of particle family produced in a hadronic collision. 
\begin{table}
	\caption{\label{tab:average-nch-ratios}Ratios of average multiplicities in different $\eta$ ranges for distributions normalized to non-single-diffractive events as measured by ALICE \cite{Adam:2015gka}}
	\vskip 2ex
	\begin{ruledtabular}
		\begin{tabular}{>{\bfseries}ccc}
			\multirow{2}{1cm}{\centering \textbf{$\sqrt{s}$ (\TeV\xspace)}} & \textbf{$\avg{n}_{|\eta|<}$} & \multirow{2}{2.5cm}{\centering \textbf{Ratio}\footnote{it should be noted that these ratios are very similar to just the ratios of $\Delta\eta$ ranges indicating that pseudorapidity density is rather flat in central region.}} \\
			{}	& \textbf{0.5 : 1 : 1.5}	& {} \\
			\midrule
			0.9		& 3.8 : 7.8 : 11.8	& 1 : 2.05 : 3.11 \\
			2.76	& 4.6 : 9.4 : 14.2	& 1 : 2.04 : 3.09 \\
			7		& 5.7 : 11.6 : 17.5	& 1 : 2.04 : 3.07 \\
			8		& 5.8 : 11.9 : 17.8 & 1 : 2.05 : 3.07
		\end{tabular}
	\end{ruledtabular}
\end{table}

Additionally, we will not limit the total charged particles number to even numbers, as is required from electric charge conservation, as this particular model is only aimed to describe the bulk behavior of the multiplicity distribution. 
In particular, it doesn't describe values at restricted $n=0$ and in first few bins. 
This choice is related to the \emph{leading particle} effect, in the sense that we describe the particle production excluding the leading particle (or particles) and thus the \enquote{total} charged-particle multiplicity $N$ can take odd values.

\paragraph{Application to ALICE measurements.}%
\begin{table*}
	\caption{\label{tab:full-model-fit}Fit parameters and $\chi^2$ for the model fit with full phase space multiplicity distribution as a weighted sum of two NBDs (presented in \cref{fig:full-model-fit-low,fig:full-model-fit-high}) and with weighted sum of NBD and Gamma distribution. }
	\vskip 2ex
	\begin{ruledtabular}
		\begin{tabular}{>{\bfseries}ccccccccccc}
			\multicolumn{11}{c}{\textbf{NBD+NBD}} \\
			\midrule[2pt]
			$\sqrt{s}$ (\TeV\xspace)	& \textbf{$\alpha$}	& \textbf{$\avg{N}_1$}	& \textbf{$k_1$}	& \textbf{$\avg{N}_2$}	& \textbf{$k_2$}	& \textbf{$A_1$}	& \textbf{$A_2$}	& \textbf{$A_3$}	& \textbf{$\slfrac{\chi^2}{\mathrm{NDF}}$}		& \textbf{Ratio $A_1$ : $A_2$ : $A_3$} \\
			\midrule
0.9     &    0.55       &   23.92       &    3.23       &   59.29       &    3.79       &   0.103       &   0.204       &   0.301       & 81.517 / 160  &   1.000 :   1.986 :   2.941 \\
2.76    &    0.59       &   31.16       &    2.55       &   87.80       &    3.69       &   0.092       &   0.182       &   0.268       & 116.085 / 230 &   1.000 :   1.979 :   2.907 \\
7       &    0.66       &   41.49       &    1.91       &  132.63       &    4.06       &   0.084       &   0.167       &   0.249       & 212.602 / 328 &   1.000 :   1.990 :   2.965 \\
8       &    0.68       &   45.35       &    1.89       &  143.20       &    4.47       &   0.085       &   0.168       &   0.247       & 259.175 / 314 &   1.000 :   1.980 :   2.910 \\
			\midrule
			\multicolumn{11}{c}{\textbf{NBD+Gamma}} \\
			\midrule[2pt]
			$\sqrt{s}$ (\TeV\xspace)	& \textbf{$\alpha$}	& \textbf{$\avg{N}_1$}	& \textbf{$k_1$}	& \textbf{$k_2$}	& \textbf{$\avg{N}_2$}	& \textbf{$A_1$}	& \textbf{$A_2$}	& \textbf{$A_3$}	& \textbf{$\slfrac{\chi^2}{\mathrm{NDF}}$}		& \textbf{Ratio $A_1$ : $A_2$ : $A_3$} \\
			\midrule
0.9     &    0.80       &   25.98       &    2.28       &    4.00       &   55.32       &   0.128       &   0.252       &   0.376       & 196.316 / 160 &   1.000 :   1.979 :   2.945 \\
2.76    &    0.70       &   30.39       &    2.50       &    4.00       &   93.34       &   0.096       &   0.189       &   0.279       & 76.353 / 230  &   1.000 :   1.972 :   2.916 \\
7       &    0.80       &   42.68       &    1.93       &    4.00       &  140.15       &   0.087       &   0.171       &   0.257       & 792.772 / 328 &   1.000 :   1.962 :   2.960 \\
8       &    0.76       &   41.84       &    1.97       &    4.00       &  141.34       &   0.088       &   0.172       &   0.256       & 270.035 / 314 &   1.000 :   1.963 :   2.920 \\
		\end{tabular}
	\end{ruledtabular}
\end{table*}%
From the known behavior of NBD we can conclude that instrumental bin $n=0$ can not be described by this procedure. 
Let us consider weighted sum model for $P_{\mathrm{tot}}(N)$ (\cref{eq:weighted-sum}) with NBDs. 
Since the parameter $k$ is unaffected by binomial reduction, the distinct maximum from the first term will be present in the $P_{\mathrm{mod}}$ and thus value of $P_{\mathrm{mod}}(0)$ will be lower than for the $n>0$ around the maximum.

However, the ALICE results \cite{Adam:2015gka,Zaccolo:2015udc} indicate an \enquote{S} shape at the start of the distribution with the value of $P_{\mathrm{exp}}(0)$ larger than the rest of the distribution even for non-single-diffractive event sample, especially in large pseudorapidity intervals and at higher $\sqrt{s}$.
Such behavior, already seen in increasing $\chi^2$ for the direct fits of distributions in central $\eta$ region can be argued to emerge from the double-diffractive and residual single-diffractive events present in the sample.
Modeling effects from such events goes beyond the scope of this paper, however we will note here that  it can be done within the proposed framework by introducing a separate term into $P_{\mathrm{tot}}(N)$ with a specific $P(n|N,\Delta\eta)$ that is based on a different pseudorapidity density. 
A viable option is a 3$^{\mathrm{rd}}$ term in the weighted sum (\cref{eq:weighted-sum}) with low $\avg{N}$ and parameter $k \gtrsim 1$ (thus lacking the structure with a maximum) with a very low reduction factor $A$ so that most of its effect is concentrated at low multiplicity $n$.
Using measurements from different experimental collaborations which have detector sets with different diffractive mass sensitivity (such as ALICE and CMS) can provide important constraints for that purpose.
This will be considered in the follow-up analysis.

Here we concentrate on modeling the bulk distribution thus excluding bin $n=0$ from fit criterion. 
It was found that scaling factor similar to one introduced for direct modeling is not needed in this formulation\footnote{This is mostly due to the fact that for non-single-diffractive sample its value is $\approx 0.94$ that is very close to $1$}, indirectly confirming that we are modeling only a subset of events included in ALICE NSD sample.
The fit criterion, for a given energy $\sqrt{s}$ is written as follows
\begin{equation}\label{eq:fit-criterion-deltaeta}
\chi^2 = N_{\mathrm{total}} \sum\limits_{\Delta\eta} \frac{1}{N_{\mathrm{bins}} (\Delta\eta)} \chi^2_{\Delta\eta}
\end{equation}
where for each pseudorapidity range $\Delta\eta$
\begin{equation}
\chi^2_{\Delta\eta} = \sum\limits_{n=1}^{n_{\mathrm{max}}} \left( \frac{P_{\mathrm{mod}}(n,\Delta\eta) - P_{\mathrm{exp}}(n,\Delta\eta)}{\sigma_n} \right)^2
\end{equation}
and the factor
\begin{equation}
N_{\mathrm{total}} = \sum\limits_{\Delta\eta} N_{\mathrm{bins}} (\Delta\eta)
\end{equation}
is intended to restore the meaning of $\chi^2$ as a sum of $N_{\mathrm{total}}$ squared deviations from normally distributed random variables.
The weighting of criteria for separate $\eta$ ranges is intended to balance the constraints.
Without weighting the experimental distribution at largest $\Delta\eta$ will have more contribution to total $\chi^2$ due to the largest number of experimental points. However, it is also less precise due to a smaller event sample.
Unfortunately, due to highly correlated uncertainties in the experimental distribution, the simple criterion is insufficient to estimate the uncertainties on fit parameters reliably and without data in more $\eta$ intervals, the fit is rather under-constrained.

It should be noted here, that the fit criterion \cref{eq:fit-criterion-deltaeta} doesn't account for a direct correlation between parameters $A_{1,2,3}$ and the unrestricted phase space average multiplicity given by parameters $\avg{N}_{1,2}$.
In fact, any (small enough) multiplicative change simultaneously in all the $A_{1,2,3}$ parameters would immediately translate into the reversed change in $\avg{N}_{1,2}$ without affecting the fit criterion.
This additional freedom in parameter choice requires limiting the average full phase space multiplicity by some external means. 
Let us consider the extreme cases first.
Decrease in the total average multiplicity yields a fit compatible with data until parameter $A$ for the largest pseudorapidity interval reaches limiting value of 1 and thus the further decrease in average unrestricted multiplicity will make the description of the largest $\eta$ range inadequate.
Average multiplicity increase is not limited by this particular method, however it is rather obvious that there are physical limitations on how many particles can be produced in a hadron collision at certain energy.
It is possible to estimate the upper limit, though it is quite large for the \TeV-scale collider. 
\begin{figure}%
	\centering%
	\includegraphics{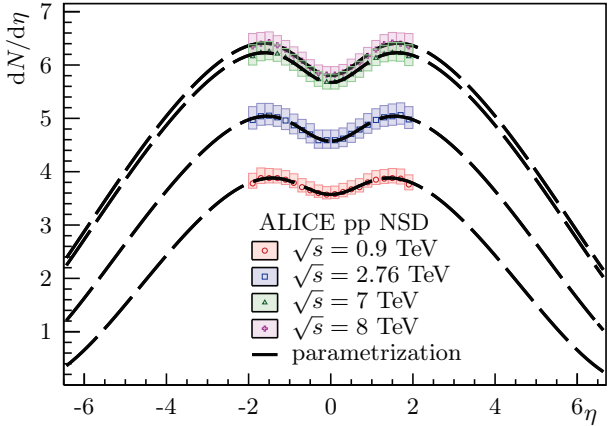}
	\caption{\label{fig:dndeta-par}Parametrization of $\slfrac{\mathrm{d}N}{\mathrm{d}\eta}$ distributions from ALICE with eqs. (\ref{eq:dndy-initial}--\ref{eq:y-to-eta}) to extract initial values for $A_{1,2,3}$.}%
\end{figure}%

This problem can be solved by placing reasonable restrictions on parameters $A_{1,2,3}$ values. 
By definition $A_{1,2,3}$ are determined by (assumed) universal pseudorapidity density functional form asymptotic behavior. 
In this publication, we estimate these parameters through parametrized $\slfrac{\mathrm{d}N}{\mathrm{d}\eta}$ distributions (see \cref{fig:dndeta-par}) fitted to ALICE data from the same dataset \cite{Adam:2015gka}. The initial values of parameters are presented in \cref{tab:A-estimate}.

We start with the simplest functional form for rapidity density
\begin{equation}\label{eq:dndy-initial}
\frac{\mathrm{d}N}{\mathrm{d}y} = \left.\avg{N}\right|_{\eta=0}\left(1 - \frac{y}{y_{\mathrm{max}}} \right)^{2k}\left(1 + \frac{y}{y_{\mathrm{max}}}\right)^{2k}\!\!\!,
\end{equation}
where $y_{\mathrm{max}} =\ln \frac{\sqrt{s}}{m_{\mathrm{p}}}$ ($m_{\mathrm{p}}$ is a proton mass), ${k = 1}$ and $\left.\avg{N}\right|_{\eta=0}$ is the particle density at $\eta=0$.
Note that this particular type of function unambiguously arises from triple pole pomeron model \cite{Alkin:2009ds,Pauk:2009zz}.
For a given particle with mass $m$ and transverse momentum $p_{\bot}$, we can relate rapidity density with pseudorapidity density as
\begin{equation}\label{eq:jacobian}
\frac{\mathrm{d}N}{\mathrm{d}\eta} = \frac{1}{\sqrt{1 + \frac{1}{b^2\cosh^2 \eta}}} \frac{\mathrm{d}N}{\mathrm{d}y}
\end{equation}
with $b = \slfrac{p_{\bot}}{m}$ and
\begin{equation}\label{eq:y-to-eta}
y = \ln \left( \frac{\sqrt{1+b^2\cosh^2\eta} + b\sinh\eta}{\sqrt{1+b^2}}\right).
\end{equation}
Using an \enquote{effective} $b$ as a free parameter, we can easily fit pseudorapidity density data at all energies with $b \approx 0.6$.
The extrapolation should be considered with caution, as it completely neglects the non-perturbative (in particular --- diffractive) contributions. 
The apparent low value of $b$ is also interesting, but it emerges as a non-trivial convolution of both $\pt$ and mass spectrum of produced particles and thus doesn't have a straightforward interpretation.

Note that the estimated ratios are quite close to those observed in average multiplicities, but estimated $A_3/A_1$ is increasing, while $\avg{n}_3/\avg{n}_1$ is decreasing with energy, indicating widening of the pseudorapidity density shape that is not seen in the simple parametrization. 
This analysis will be redone as soon as new measurements (in particular multiplicity distributions in additional pseudorapidity intervals) will become available.
\begin{figure*}%
	\centering%
	\resizebox{16.7cm}{!}{
	\subfloat{%
		\includegraphics{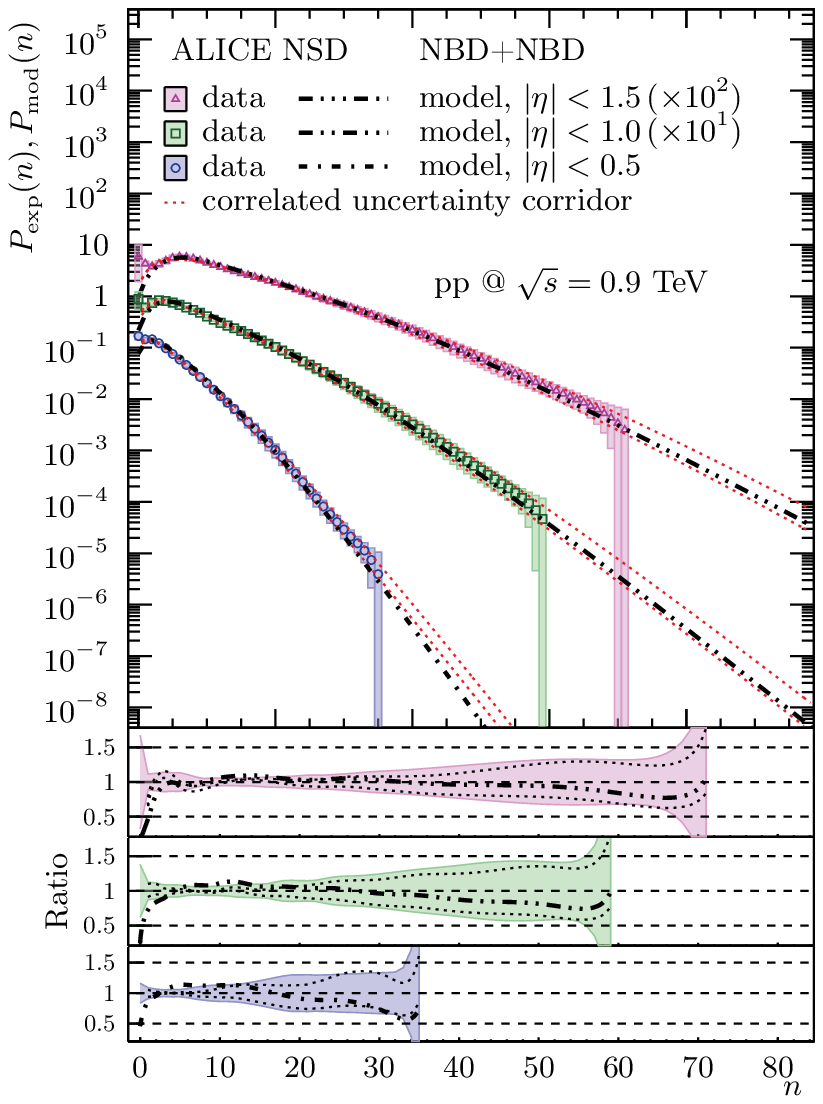}
	}%
	\subfloat{%
		\includegraphics{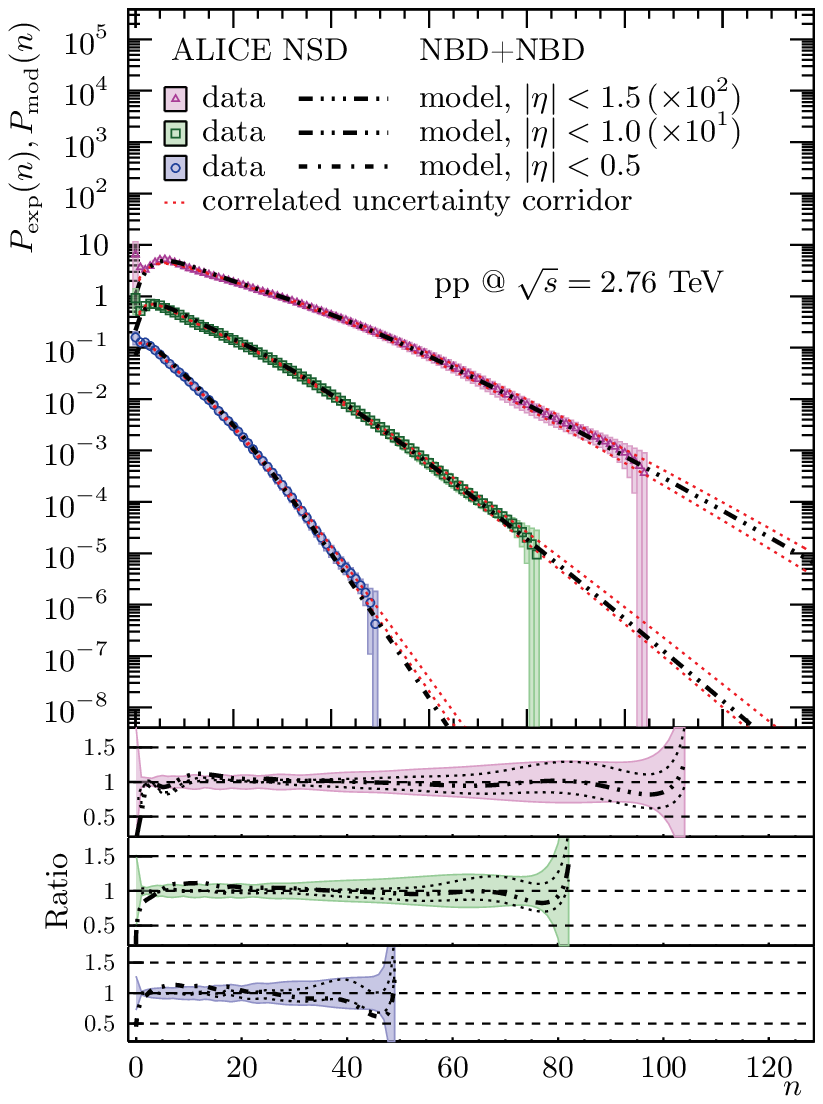}
	}%
	}
	\caption{\label{fig:full-model-fit-low}A model fit of ALICE non-single-diffractive multiplicity distributions in proton-proton collisions with weighted NBD sum model for a full phase space distribution and binomial reduction at $\sqrt{s}$ = 0.9 (left) and 2.76 (right)~\TeV. Bounding direct double-NBD fits from ALICE are added to indicate a correlated uncertainty corridor. Curves at different $\eta$ intervals are displaced vertically for clarity. Shaded areas indicate combined statistical and systematic uncertainty of experimental data. Ratios of model fits to experimental data are presented in the bottom of the figure with the same convention for the shaded area.}%
\end{figure*}%

\begin{figure*}%
	\centering%
	\resizebox{16.7cm}{!}{
	\subfloat{%
		\includegraphics{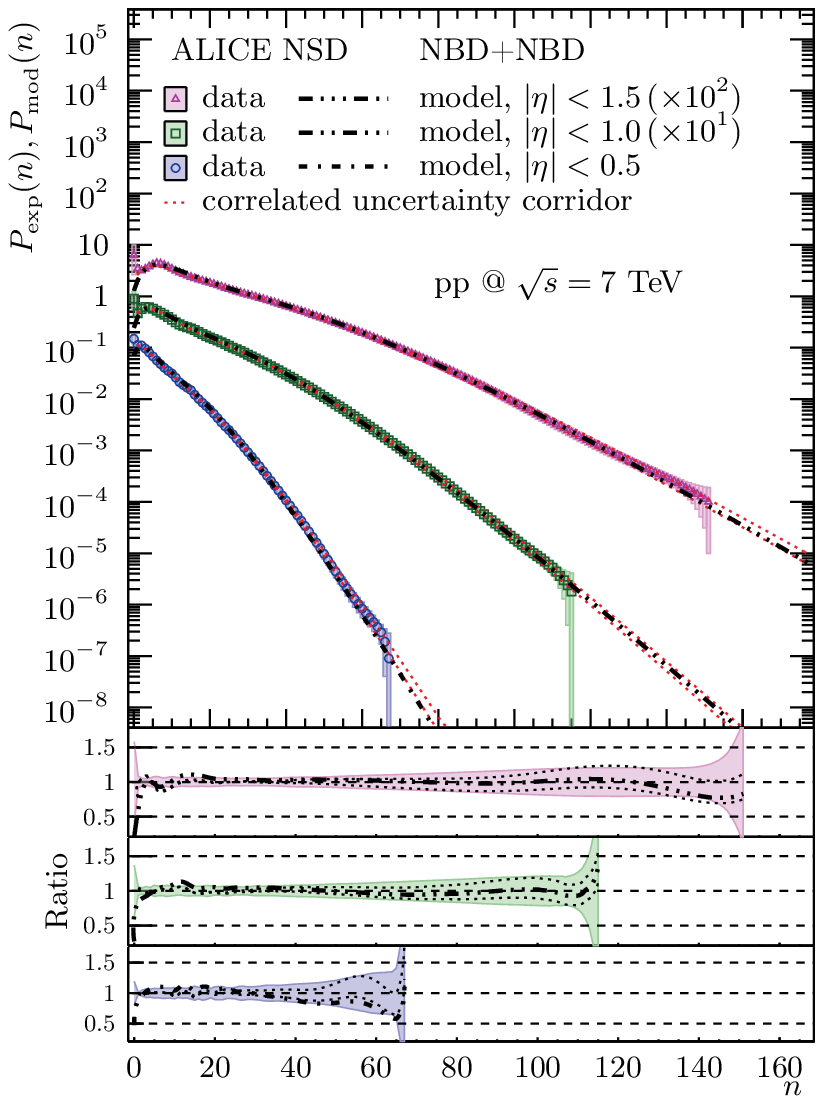}
	}%
	\subfloat{%
		\includegraphics{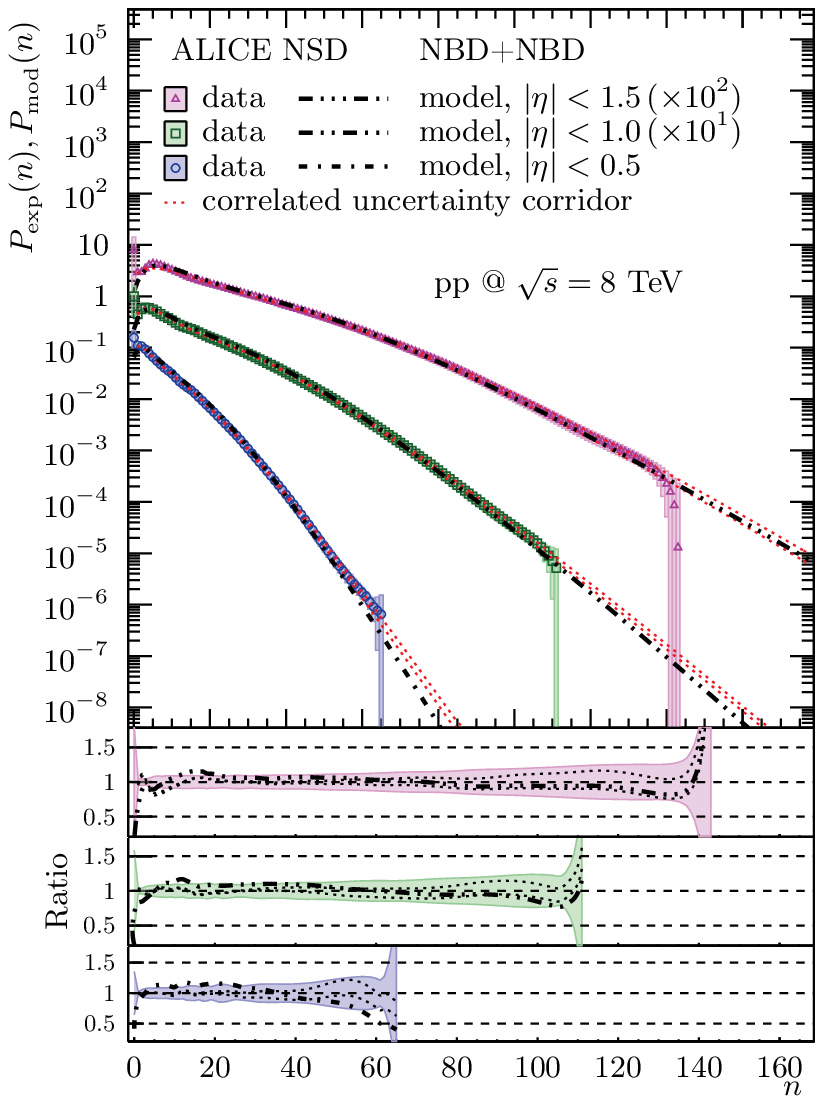}
	}%
	}
	\caption{\label{fig:full-model-fit-high}A model fit of ALICE non-single-diffractive multiplicity distributions in proton-proton collisions with weighted NBD sum model for a full phase space distribution and binomial reduction at $\sqrt{s}$ = 7 (left) and 8 (right)~\TeV. Bounding direct double-NBD fits from ALICE are added to indicate a correlated uncertainty corridor. Curves at different $\eta$ intervals are displaced vertically for clarity. Shaded areas indicate combined statistical and systematic uncertainty of experimental data. Ratios of model fits to experimental data are presented in the bottom of the figure with the same convention for the shaded area.}%
\end{figure*}%

\begin{table}
	\caption{\label{tab:A-estimate}Estimated values for $A_{1,2,3}$ from extrapolated parametrization of ALICE NSD $\slfrac{\mathrm{d}N}{\mathrm{d}\eta}$ measurements \cite{Adam:2015gka}.}
	\vskip 2ex
	\begin{ruledtabular}
		\begin{tabular}{>{\bfseries}ccc}
			{\centering \textbf{$\sqrt{s}$ (\TeV\xspace)}} & \centering \textbf{$A_{-0.5}^{+0.5}$ : $A_{-1}^{+1}$ : $A_{-1.5}^{+1.5}$} & {\centering \textbf{Ratio}}	\\
			\midrule
			0.9  & 0.104 : 0.212 : 0.323 & 1 : 2.040 : 3.112 \\
			2.76 & 0.091 : 0.186 : 0.284 & 1 : 2.044 : 3.127 \\
			7    & 0.083 : 0.170 : 0.260 & 1 : 2.041 : 3.119 \\
			8    & 0.082 : 0.168 : 0.257 & 1 : 2.043 : 3.126 \\
		\end{tabular}
	\end{ruledtabular}
\end{table}

Due to the significant correlated errors on experimental distribution, parameters other than the average multiplicity of the full phase space multiplicity distribution are largely unrestricted, and the usual fit procedure is unable to produce a reliable uncertainty estimate on fit parameters. 
Thus, we will avoid drawing any rigorous conclusions in this publication based on the parameter evolution. 
We will, however, outline the direction of theoretical and experimental inquiry that can be pursued to verify and improve the methods described here.

\begin{figure}[t]%
	\centering%
	\resizebox{8.1cm}{!}{
	\includegraphics{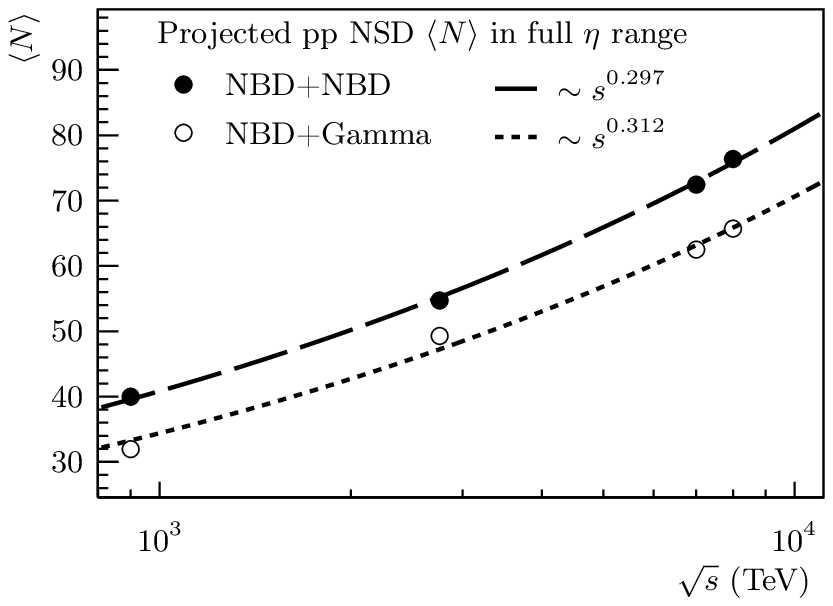}
	}%
	\caption{\label{fig:avgN-vs-sqrts}Estimated unrestricted phase space average multiplicity in proton-proton collisions as a function of center-of-mass energy $\sqrt{s}$. Powerlaw fits are added to indicate the trend.}%
\end{figure}

\begin{figure*}%
	\centering%
	\resizebox{16.7cm}{!}{
	\includegraphics{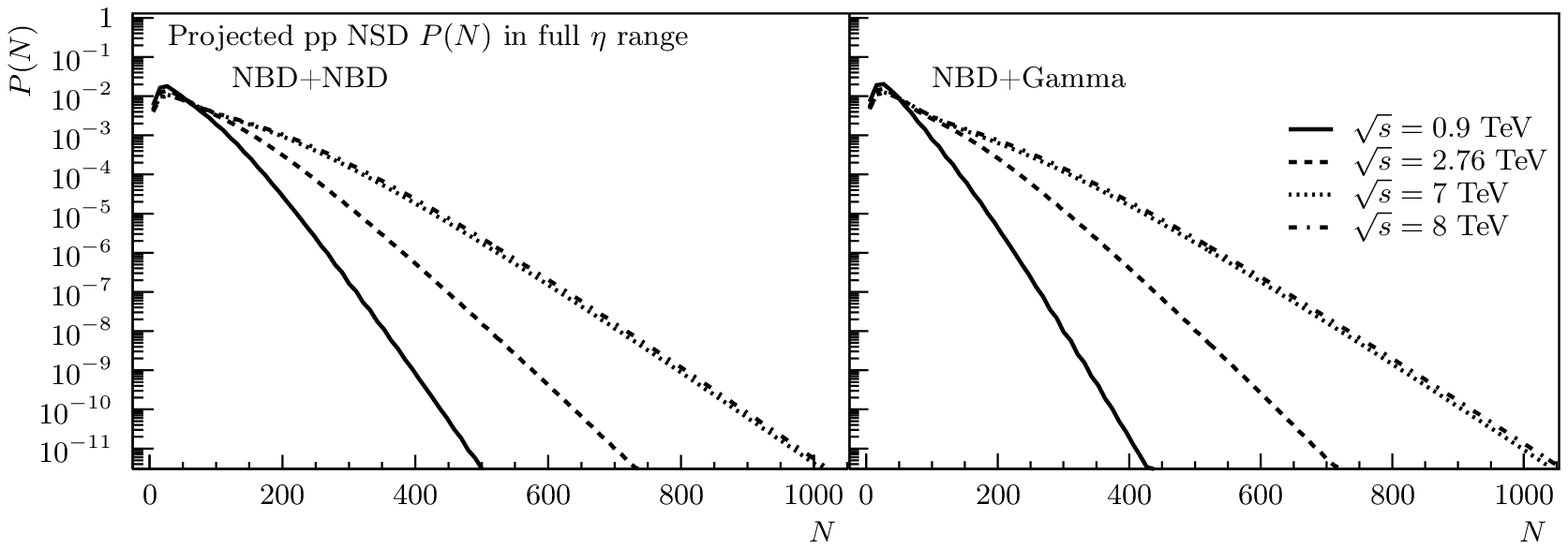}%
	}%
	\caption{\label{fig:full-MD}Unrestricted phase space multiplicity distributions for proton-proton collisions at various energies projected from NBD+NBD (left) and NBD+Gamma (right) models.}%
\end{figure*}

The resulting model fit with $P_{\mathrm{tot}}(N)$ defined by a weighted sum of two NBDs is presented in \cref{fig:full-model-fit-low,fig:full-model-fit-high} and the corresponding parameters in \cref{tab:full-model-fit}.
Weighted sum of NBD and Gamma distribution also provides simultaneous description of all three restricted phase space multiplicities (see \cref{tab:full-model-fit}), however, with lesser compatibility, especially at low center-of-mass energies. 
It also features somewhat steeper growth of the unrestricted phase space average multiplicity with $\sqrt{s}$ (see \cref{fig:avgN-vs-sqrts}), but within the expected uncertainty we may say that these trends are similar.
We note here, that due to low fit quality for NBD+Gamma option, there is considerably more freedom in determining $\avg{N}_{1,2}$ and $\alpha$ parameters, the apparent trend may change. 
However, the result for NBD+NBD is more reliable and stable. 
Overall unrestricted multiplicity distributions for energies from 0.9 to 8~\TeV\xspace extrapolated from NBD+NBD and NBD+Gamma models are presented in \cref{fig:full-MD}.

\subsection{Discussion of results}
[ht]%
First of all, let us reiterate once again the fact that due to experimental limitations, specifically the correlated uncertainties, we are unable to extract errors on model parameters without more data and additional restrictions from different observables. 
In particular, we cannot draw rigorous conclusions from the trends of shape parameters $k$ for both NB and Gamma distributions, as those may not hold after repeating this analysis with more data, however, we will discuss the possible physical interpretations of the observed behavior.

The restriction factors $A_{1,2,3}$ differ between \text{double-NBD} and NBD+Gamma models. 
In both cases they decrease with energy as well as their ratios, however, the ratios differ both from those estimated by pseudorapidity density parametrization and the ratios of average multiplicities.
Note that due to the fact that low-multiplicity part of the distribution is not fully modeled, the projected average multiplicity is slightly higer which may explain the apparent difference in ratios. 
It also can confirm the observation from parametrized pseudorapidity density, that its shape widens with the growth of $\sqrt{s}$ and may be wider than our parametrization suggests.
Alternatively it can indicate the inadequacy of shape universality assumption, that can be tested using data from wider pseudorapidity intervals.

Finally, the possibility of internal inconsistency between ALICE multiplicity measurements at different energies cannot be discarded. 
In particular, the direct fits, both from ALICE and performed here, produce notably different parameters (specifically, the weight $\alpha$ and shape parameters $k_{1,2}$) between measurements at 7 and 8~\TeV.
As there is no expectation for measurements being considerably different with just 14\% increase in center-of-mass energy (and completely negligible on $\ln s$ scale of high-energy physics), such properties can be both explained with the internal inconsistency and the appearance of new scattering modes at these energies. 
Note that the stability of Gamma component $k$ parameter, observed for direct fits, still holds for the combined model, though with lesser compatibility to data. 
This may change when the analysis is performed with more pseudorapidity intervals when they become available.
Regardless, such stability is an indication that a sub-sample of events, governed by a particular particle production mechanism, exhibits scaling behavior already at \TeV\xspace energies. 
More detailed analysis is required, however we will again point out that the most significant problem of multiplicity measurements is correlated uncertainty.
The lack of direct information on correlations removes a number of statistical tools at our disposal, that can be used to extract physical results in a straightforward way.

Note here, that we are not modeling the sub-sample of events that gives rise to \enquote{S} shape at low multiplicities that becomes more pronounced with increasing $\eta$ range, which obviously has a different shape of pseudorapidity density.
Inclusion of additional terms in $P_{\mathrm{tot}}(N)$ at this stage would not improve the analysis, however, it may be required to describe wider $\eta$ intervals. 

A possible expansion of the method suggested here would include simultaneous modeling of both the pseudorapidity density and unrestricted phase space multiplicity distribution, ultimately without the factorization assumption used to construct $P(n|N,\Delta\eta)$.

%% file: conclusion.tex
Phenomenological model, presented in this paper, is able to successfully simultaneously describe several multiplicity distributions in varying pseudorapidity intervals at a given energy. 
Using experimental distributions in more pseudorapidity intervals this can be improved further by constraining weight parameter $\alpha$ that controls the sample separation, and the shapes of separate distributions of the either sample controlled by parameters $k_{1,2}$.
Such approach paves a way to making charged-particle multiplicity distributions more useful as a tool to probe, in particular, the balance between soft- and hard-QCD processes in collider experiments, as it leaves less free parameters to describe phenomenologically or theoretically.
As demonstrated by the analysis performed, such unified approach already allows better sample separation than direct fitting of restricted phase space multiplicity distributions' shapes, though still limited by the unknown correlations in experimental systematic uncertainty. 
An important consequence of this is the potential for developing experimental techniques to measure overall multiplicity distributions as well as an additional sample separation criteria for proton-proton scattering that lack the clear analogue of ion-ion physics centrality.

It is shown, that charged particle multiplicity distributions at energies ranging from 0.9 to 8 \TeV\xspace are well-described by the model that assumes the uncorrelated mix of two event samples, with different multiplicity distribution shapes, both given by negative-binomial distribution. 
Gamma distribution, which can be regarded as scaling limit of NBD, is also compatible with the second sample, at least at higher energies, suggesting that scaling behavior may be manifesting in different event samples at different $\sqrt{s}$, explaining the apparent lack of the overall scaling. 
It can be suggested, that with increase in center-of-mass energy, and thus the increase in average multiplicity, leading to $\slfrac{\avg{n}}{k} \gg 1$, both components will become compatible with the Gamma distribution reaching the corresponding scaling limits. 

The shape of particle pseudorapidity density distribution plays an important role within this framework. 
If the experimental results for pseudorapidity density in proton-proton collisions, classified somehow in bins by the overall event multiplicity (or related inelasticity measure) become available, it will be possible to refine this model possibly including the dependence of restriction factors $A$ on total number of particles $N$.
However, even the usual averaged pseudorapidity distribution measured in wider $\eta$ interval can be used to extract more precise values of $A$ thus constraining the rest of model parameters.

One of the important consequences of the proposed model is the stability of the NBD shape parameter $k$ across different pseudorapidity intervals which contradicts a long-standing assumption, based on the fact that direct fits yield different values for different $\eta$ ranges.
Moreover, this parameter, through its relation to the dispersion of NBD, is also directly related to two-particle correlation function \cite{Giovannini:1985mz} thus this observation potentially has significant theoretical consequences.

Overall, the described approach offers a novel view of restricted phase space charged particle multiplicity distributions that has a lot of potential. 
This model will be applied to upcoming experimental data and we will continue our work on refining different aspects of it, in particular, by combining more sources of information about particle production in proton-proton collisions.